# The Damage to Lunar Orbiting Spacecraft Caused by the Ejecta of Lunar Landers

Philip T. Metzger[1] and James G. Mantovani[2]

[1] Florida Space Institute, University of Central Florida, 12354 Research Parkway, Suite 214, Orlando, FL 32826; PH (407) 823-5540; email: philip.metzger@ucf.edu
[2] Swamp Works, Exploration Systems & Development, Mail Code UB-E-2, NASA Kennedy Space Center, FL 32899; (321) 867-1870; james.g.mantovani@nasa.gov

## ABSTRACT

This manuscript analyzes lunar lander soil erosion models and trajectory models to calculate how much damage will occur to spacecraft orbiting in the vicinity of the Moon. The soil erosion models have considerable uncertainty due to gaps in our understanding of the basic physics. The results for ~40 t landers show that the Lunar Orbital Gateway will be impacted by 1000s to 10,000s of particles per square meter but the particle sizes are very small and the impact velocity is low so the damage will be slight. However, a spacecraft in Low Lunar Orbit that happens to pass through the ejecta sheet will sustain extensive damage with hundreds of millions of impacts per square meter: although they are small, they are in the hypervelocity regime, and exposed glass on the spacecraft will sustain spallation over 4% of its surface.

## INTRODUCTION

The Moon is a reduced-gravity, airless body, so a rocket landing on the Moon potentially blows ejecta to higher than orbital altitudes or even completely off the Moon. To plan lunar missions, it is important to understand the trajectories and flux of the ejecta to protect orbiting spacecraft. Protection may be provided through timing the landings relative to positions of the orbiting spacecraft (COLlision Avoidance, or COLA), and through the construction or deployment of landing pads. Mitigation may be costly or sometimes impossible, so we need requirements derived from science on how much damage will occur without mitigation.

## RISKS DUE TO UNKNOWNS IN THE PHYSICS

The science of interactions between rocket engines and planetary regolith is still very immature and large gaps exist in our understanding, including the following two areas.

**Cratering Regime Physics.** Prior worked showed that there are various *regimes* of interaction, and different regimes will be "turned on" depending on the conditions of the rocket exhaust flow, the conditions of the regolith, and the conditions of the planetary environment (mainly the atmosphere and gravity). As near as we can tell, lunar landings in the Apollo program and smaller robotic missions only "turned on" the surface erosion regime, where gas flow traveling horizontally across the surface of the soil entrains individual particles. Fig. 1 shows the terrain beneath the Apollo 11 Lunar Module after landing. It was swept clean but there was no central crater. Other regimes have been seen in experimental work. They include: bearing capacity failure, diffusion-driven flow, diffused gas eruption, and shock-induced fluidization. Not enough work has been done to map the parameter space where these regimes "turn on", so it is unclear whether or when a larger lunar lander could induce any of them. Those regimes have the potential to create a deep crater under the lander, and that would redirect the gas flow sending ejecta into a more upward direction. Also, they may induce a higher entrainment rate into the gas flow. Until more work is done, it is impossible to know if that will occur. The present work assumes only surface erosion will occur, so this is a risk we are carrying forward.

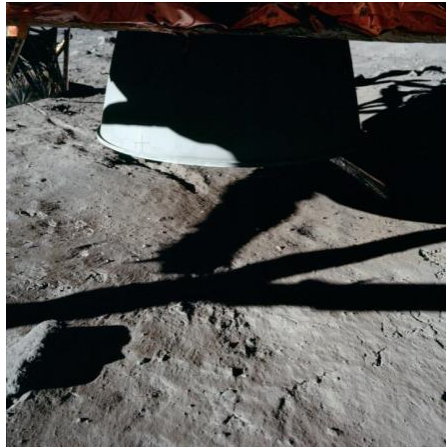

Figure 1. No crater under the Apollo Lunar Modules after landing.

**Erosion Rate Physics.** Prior work studied the erosion rate (or emission rate or entrainment rate) that occurs in the surface erosion regime in lunar landings. Experiments were performed at ambient atmospheric pressure, varying the gas jet parameters (molecular weight, exit velocity, jet diameter, height of jet exit plane), the sand parameters (density, particle size), and gravity (in reduced gravity aircraft experiments). The experimental work showed that erosion rate in continuum conditions scales proportionately to the momentum flux of the gas to the unity power, and inversely to the potential energy in the sand (grain density multiplied by grain diameter multiplied by gravity) (Metzger et al., 2010). In other words, it was the square of a densimetric Froude number. This made physical sense, so it was accepted as probably correct despite some gaps in the measurements. However, when the erosion rate was measured from the optical density of dust in the Apollo landings where the gas was rarefied, the erosion rate was found to scale as local shear stress to the 2.5 power (Lane and Metzger, 2015). Momentum flux to the unity power does not scale the same as local shear stress to the 2.5 power. Other experiments were performed in a large vacuum chamber and it was found that, keeping momentum flux constant, the erosion rate begins to increase when the Knudsen number relative to a sand grain radius is between 0.01 and 0.1, that is, the transitional flow regime (Metzger, 2015). This is shown in Fig. 2. It was beyond the capability of the vacuum chamber to test in the regime with Knudsen numbers larger than 0.1, so the full curve into the region of the Apollo landings could not be obtained. A sand grain radius is comparable to the roughness length scale of the surface of the material, which determines the velocity profile in the boundary layer. This result suggests that diffusion of momentum through the boundary layer becomes more efficient as the gas becomes rarefied and may potentially explain the different scaling in rarefied versus continuum conditions. The simplest model is to take the above results at face value. A hyperbolic tangent function fits the transitional flow regime data in the vacuum chamber experiments, so the model uses the hyperbolic tangent to interpolate between the continuum scaling (terrestrial measurements) and rarefied scaling (Apollo landing measurements).  There is a significant risk that this is incorrect, either because there are additional transitions in the physics that have not been identified, or because the methodology of measuring erosion rate in either regime has not accounted for all dependencies such as the role of turbulence. Because we have no better model to estimate erosion rate in these extreme conditions, this is a second risk that we are carrying forward.

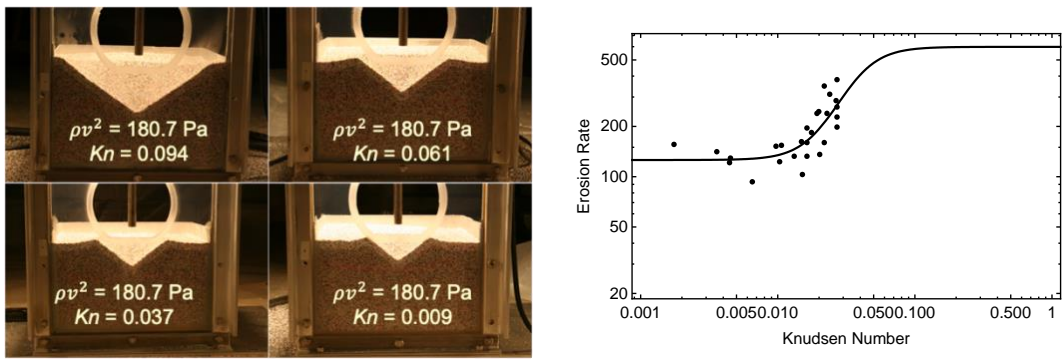

**Figure 2. Left: Four vacuum chamber experiments showing erosion rate scaling with *Kn*. Right: Hyperbolic tangent fitted to many vacuum chamber tests.**

### EROSION RATE MODEL

Terrestrial experiments to derive the erosion equation included small-scale, vertical jet impingement experiments such as shown in Figure 2 (Right). The cases that were closest to the relevant lunar case were the ones performed in the large vacuum chamber using lunar soil simulant JSC-1A. To analyze these tests, the small scale experiments of Rajaratnam and Mazurek (2005) were used. The data show the shear stress as a function of distance from an impinging jet. The data from their cases 4 and 8 were replicated in Fig. 3. Both cases have jet exit height $H$ = 152.4 mm and jet exit diameter $D$ = 12.7 mm.  Case 4 has jet exit velocity $U_0$ = 60.96 m/s. Case 8 has $U_0$ = 68.58 m/s. Case 4 has exit Reynolds number $Re_0$ = 51271. Case 8 has $Re_0$ = 57680. Case 4 has peak shear stress $\tau_{0,max}$ = 20.01 Pa so normalized by total momentum flux at the jet exit it is $\tau_{0,max}/(\rho_0 U_0^2 \pi D^2)$ = 0.0045. Case 8 has $\tau_{0,max}$= 20.49 Pa so normalized to total momentum flux at the jet exit $\tau_{0,max}/(\rho_0 U_0^2 \pi D^2)$ = 0.0036. The biggest difference between the two cases is surface roughness. For Case 4, surface roughness is ks = 1.73 mm. For Case 8, ks = 8.23 mm.

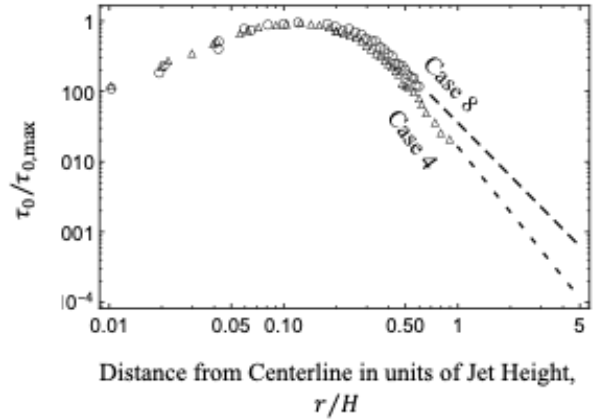

Figure 3. Normalized Shear Stress of Rajaratnam and Mazurek (2005), Fig. 3.

Table 1. JSC-1A Test Results.

| Case | $Re_0$ | $Kn_0$ | $U_f$ (m/s) | No Erosion | | With Erosion | | F (kg/s/m²/Pa) |
|---|---|---|---|---|---|---|---|---|
| | | | | k | Flow | k | State | |
| 1 | 4902 | 0.0065 | 2.69 | 5.4 | Trans. Rough | 160.8 | Fully Rough | 4.883 |
| 2 | 5413 | 0.0065 | 2.97 | 5.9 | Trans. Rough | 177.5 | Fully Rough | 3.586 |
| 3 | 6230 | 0.0065 | 3.42 | 6.8 | Trans. Rough | 204.3 | Fully Rough | 3.262 |
| 4 | 5005 | 0.0065 | 2.75 | 5.5 | Trans. Rough | 164.1 | Fully Rough | 4.211 |
| 5 | 2145 | 0.0338 | 6.13 | 2.3 | Smooth | 70.4 | Fully Rough | 5.754 |
| 6 | 1293 | 0.0986 | 10.78 | 1.4 | Smooth | 42.4 | Trans. Rough | 5.427 |
| | | | | | | | Average F value = | 4.521 |

For the experiments with JSC-1A and nitrogen gas, $D$ = 9.4 mm, $H$ = 2.54 cm, $U_0$ = 11.5, and gas density (since it was in a vacuum chamber and not at ambient), $\rho_0$ = 1.253 kg/m³. Gas exit temperature $T_0$ = 272.56 K. Finding viscosity $\nu$ on a lookup table for nitrogen, we calculate the Reynolds Number $Re_0$ = 4902.39 and Knudsen number relative to the jet exist diameter $Kn_0$ = 0.0065. Per Fig. 7 of Rajaratnam and Mazurek (2005), $\tau_{0,max} = 0.4\,\rho_0 U_0^2\, D^2/H^2$. Friction velocity is $U_f = \sqrt{\tau_{0,max}/\rho_0} = 2.69$ m/s. (Note, $\rho = \rho_0$ everywhere in the experiment since the flow is subsonic and equalized to background pressure.) Roughness of the simulant's surface is equal to the average JSC-1A grain radius = 43.9 μm. Roughness normalized by $H$ is 0.0017. With erosion, roughness is normally 30 times greater, so normalized by $H$ it becomes 0.0519. The roughness in wall units with no erosion is therefore 0.0017 $U_f \rho/\nu$ =5.4, and with erosion it is 160.8. Therefore, the flow is transitionally rough over JSC-1A when there is no erosion, but it transitions to fully rough where erosion begins. The radius of the initial erosion hole was measured at 0.576 cm. Integrating the function for shear stress in Fig. 3 over only that radius, then dividing by momentum flux= $\rho_0 U_0^2 \pi D^2$, we find that the normalized total traction operating in the scour hole. This is repeated for both curve fits of Fig. 3 and averaged. We find that the average of the total traction divided by momentum flux is $T/(dP/dt) = 7.1$ +/- 15%. The initial volumetric growth rate of the crater was obtained by curve fitting onto the crater shape for the first few timesteps in the video and was found to be 2.7 cm³/s, so converting to mass of eroded JSC-1A, accounting for its bulk density, we find the mass erosion rate $\dot{m} = 4.07$ g/s. Normalizing by the total traction (integrated shear stress) in the hole we find the erosion factor, $F$=4.883 kg/s/m²/Pa. This process is repeated for each of the relevant experiments and the results are reported in Table 1. The average of all six cases gives $F$=4.883 kg/s/m²/Pa, with an uncertainty of ~30%.

On the Moon the erosion rate will be faster due to lower gravity. Reduced gravity aircraft experiments in 1 g and 1/6 g with JSC-1A were performed by Metzger, et al. (2009). The erosion rate was 3.4 times faster at 1/6 g. This was not the expected value $g_{Earth}/g_{Moon} = 6$ that was seen in experiments with coarser granular materials. Metzger et al. (2009) attributed that difference to the cohesion of the JSC-1A which becomes more significant at lower gravities, and which is greater for JSC-1A than for coarser materials due to the finer particle sizes. Using this factor of 3.4, the erosion factor for the lunar environment is $F_{Moon} = 3.4 \times 4.521 = 15.39$ kg/s/m²/Pa. Multiplying by the local shear stress from the rocket exhaust predicts the local erosion rate where $Kn \lesssim 0.1$ (continuum flow) relative to the radius of an average soil grain,

$$\dot{m} = 15.39\ \sigma\ \text{kg s}^{-1}\ \text{m}^{-1} \qquad (1)$$

In regions where $Kn \gtrsim 1$ relative to the radius of an average soil grain, we use the relationship obtained by Lane and Metzger (2015) obtained from optical density in the Apollo 12 landing,

$$\dot{m} = 0.0222\ \sigma^{2.52}\ \text{kg s}^{-1}\ \text{m}^{-1} \qquad (2)$$

For the transitional region, $0.1 \lesssim Kn \lesssim 1$ relative to the radius of an average soil grain, we are guided by the experiments that spanned a portion of that range. To reach larger $Kn$ within the capability of the vacuum chamber we needed coarser sand grains, so a coarse quartz sand was used. The results are shown in Fig. 2 (Right). It is fit well by a hyperbolic tangent function. The upper value of the hyperbolic tangent is not constrained by the test data but will be constrained by Eq. (2) when constructing the full theory. The lower value of this hyperbolic function was constrained by the $F$ value for quartz sand in continuum conditions in this example, but will be constrained by Eq. (1) using JSC-1A when constructing the full theory.

The full erosion rate theory uses the best fit parameters for the hyperbolic tangent in Fig. 2 (Right) except replacing the asymptotic erosion values on the left and right sides to match Eqs. (1) and (2), respectively. However, when Eq. (2) was developed, the values of $\sigma$ were obtained by computational fluid dynamics simulations. Here, we used Roberts' Equations (Roberts, 1964) to predict the laminar shear stress everywhere around the lander. Applying the erosion rate theory in this way it predicts the erosion too high by about 50%, so we modified the coefficient in Eq. (2) from 0.0222 to 0.03119 so the theory applied to the Lunar Module using Roberts' equations predicts the same 2.6 tons total erosion as measured by Lane and Metzger (2015). The result is shown in Fig. 4. This theory has obvious weaknesses but it is the best available.

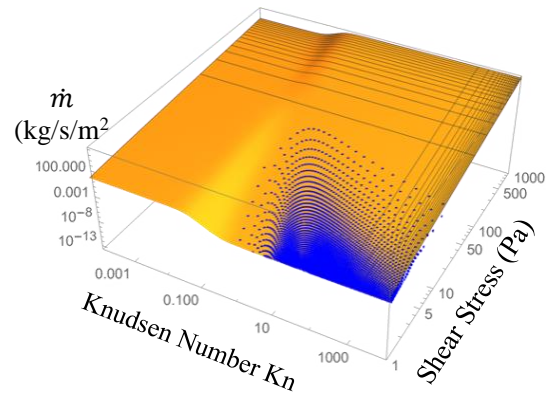

Figure 4. Erosion rate model. Blue points represent where the Apollo LM was operating.

**EJECTA TRAJECTORIES**

Prior work by Lane and Metzger (2012) showed the complex relationships of angles and speeds that particles of different sizes travel when blown by a lunar lander from a flat surface. For the majority of the landing, the majority of the ejecta travels in a broadly conical sheet about 1-3 degrees above the local horizontal (i.e., an ejecta cone with half angle 87-89 degrees). This was also seen when measuring the shadows of the lunar module on the ejecta cone (Immer et al., 2011). For the purposes of the present study, the particle trajectories will be simplified by assuming they are all in a cone of 1-3 degrees above local horizontal, uniformly distributed within that range of angles. For long distances from the Moon, Kepler's equation was used to plot the paths of the particles. An example is shown in Fig. 5 (Left). For short distances, we approximate using straight line trajectories in the cone described above.

The velocities of the ejecta are adapted from Lane and Metzger (2012). This part of the calculation assumed a 41.3 t lander (67 kN thrust) with a single engine to define the flow field. Here we assume it thrusts until 30 cm above the surface to calculate worst-case ejecta velocities. The maximum ejecta velocity for the finest dust equals the exit velocity of that propellant, which is about 4,500 m/s assuming hydrogen/oxygen (compared to about 3,100 m/s for the Apollo LM, which used Aerozine and Nitrogen Tetroxide). Particles ejected from closer to the centerline of the vehicle generally get accelerated to faster velocities before they "run out into vacuum", and smaller particles generally go faster than larger ones since drag force to inertial force scales $\sim 1/d$ where $d$ is particle diameter. In the modeling for Fig. 5 (Left), we crudely assumed the particles of each size $d$ are uniformly distributed in velocity between 30% and 100% of the maximum velocity shown in Fig. 5 (Right). The worst-case ejecta flux occurs when the vehicle is closest to the ground, and at that time the vast majority of erosion derives from very close to the vehicle so we will further simplify that all the ejecta travel at the maximum velocity shown in Fig. 5 (Right).

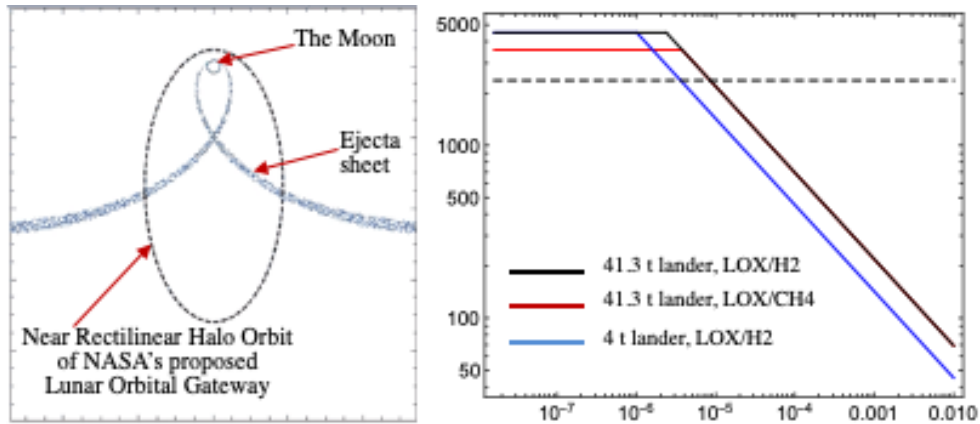

**Figure 5. Left: Simulation of ejecta sheet traveling far from the Moon. Right: Maximum velocities of each size ejecta for this case with estimated comparisons.**

**DAMAGE PREDICTIONS**

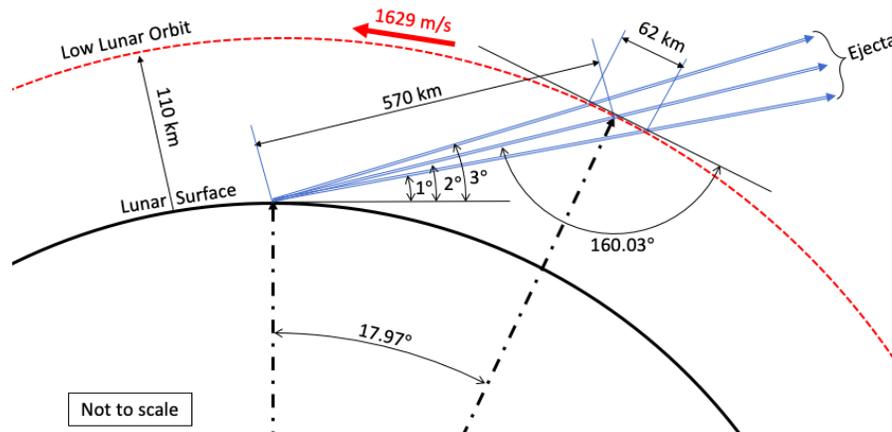

**Figure 6. Geometry of Ejecta Flux Impacting a Spacecraft in Low Lunar Orbit**

Applying this model to the case of NASA's proposed Lunar Gateway and a 40 t lunar lander (one engine) descending at 1 m/s with engine cutoff 1.5 m above the lunar surface, it predicted order-of-magnitude 2,000 to 10,000 particle impacts per square meter each time NASA's Gateway flies through the ejecta sheet, which will probably be several times before the sheet is dispersed. Work is ongoing to quantify its dispersal. The particles at that altitude are mainly 10 $\mu$m and smaller. The impact velocity will be only in the range of a few hundred meters per second since the Gateway's orbit is slow and the particles have decelerated against gravity by the time they reach that altitude. Therefore, the impacts are sub-hypervelocity. Scaling according to Sheldon and Kanhere (1972) shows that the volume of an impact pit from a sub-hypervelocity impact scales with the cube of the impact velocity. Experimental data presented by Ruff and Wiederhorn (1979) includes the amount of eroded target aluminum from particulate impacts up to 450 ft/s, so scaling to 300 m/s we calculate the divot caused by each impacting particle will be order-of-magnitude the same volume as the impacting particle. Assuming worst case, if Gateway passes through the ejecta sheet 10 times and receives 10,000 impacts/m² each time, then after 100 landings only 0.08% of its surface will be abraded a few microns deep and 99.92% of the surface will not be affected. This is important for the environmental specification but is not too severe.

Applying this to spacecraft in low lunar orbit being damaged by a 40 t lander (but with four widely spaced engines, so ejecta flux in most directions of is similar to a single engine 10 t lander), we choose a test case of $H=110$ km altitude to match the parking orbit of the Apollo Command and Service Module. At this altitude, the orbital velocity is $v_{LLO}$ = 1629 m/s. Ejecta leaving the lunar surface with a local angle of 2 degrees and traveling an approximately straight line will travel $R = 570$ km before reaching that altitude. Analyzing the trigonometry per Fig. 6, the distance the spacecraft will travel through the ejecta sheet is 62 km, so the passage time will be $\Delta t$ =11.5 ms. The relative velocity of ejecta hitting the spacecraft is

$$v_{\text{relative}} = \cos(180° - 160.03°) \, v_{\text{ejecta}} + v_{LLO} \qquad (3)$$
$$= 0.940 \, v_{\text{ejecta}} + 1{,}629 \text{ m/s}$$

Ejecta traveling 4,500 m/s will hit the spacecraft at $v_{\text{relative}}$ = 5,858 m/s. This is well into the hypervelocity regime. Due to the relative velocity, the flux hitting the spacecraft will also be increased relative to the source,

$$\psi_{\text{impact}} = \frac{\dot{M}}{2 \pi R^2 \Delta\theta} \int_0^{d_{\max}} P(d) \frac{0.940 \, v(d) + v_{LLO}}{v(d)} \, dd$$
(4)

Where $\dot{M} =$ is the total rate ejecta is liberated from the landing site (integrating $\dot{m}$ over the shear stress of the gas around the lander), $P(d) =$ the mass-weighted particle size distribution of lunar soil, $\Delta\theta = 2° = 0.035$ rad, and $v(d)$ is the particle velocity per Fig. 5 (Right). $d_{\max} =$ the largest particle size that reaches Low Lunar Orbit altitude (i.e. has an elliptical trajectory with apoapsis of the lunar radius plus 110 km), which according to orbital dynamics is a particle that is ejected with initial velocity of 1,698 m/s. Per the equation for Fig. 5 (Right), that is $d_{\max} = 16.7 \, \mu m$.

Using Roberts' equations for a 40 t, four-engine (widely spaced engines) lander to calculate $\sigma$ and $Kn$ around the lander at various heights above the surface and using those as input to the erosion rate equation of Fig. 4, then integrating the erosion rate around the lander at each height of the vehicle, we find $\dot{M}$ as a function of height as shown in Fig. 7 (Left). The calculation assumes each engine supports ¼ the weight of the vehicle and the engines are far enough apart that the maximum erosion is close to the engine, not on the plume reflection planes between engines, so engine interactions are ignored in this approximation.

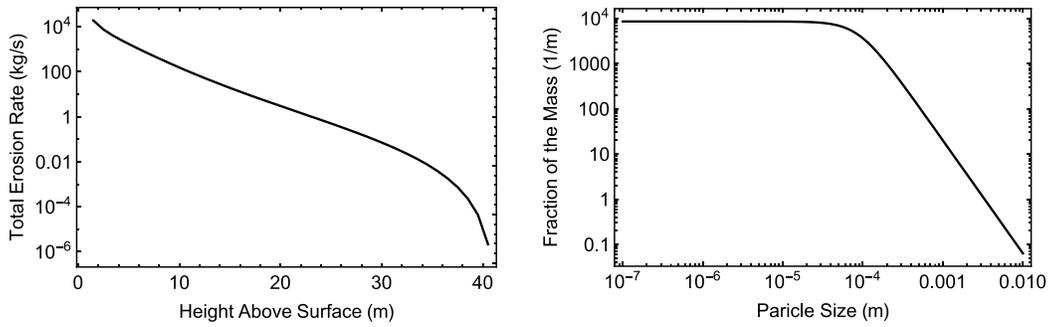

**Figure 7. Left: Erosion Rate as a function of height above the surface. Right: Particle Size Distribution Model.**

The erosion rate when the engine is 1.5 m above the surface is 19,238 kg/s. This is high but reasonable based on comparison with Apollo and many other simulations and experiments. The particle size distribution model in Fig. 7 (Right) is

$$P(d) = \frac{C_1}{11.11 + 0.0015\, d^{2.5}} \tag{5}$$

where $C_1$ is for normalization such that the integral over all particle sizes is unity for $0 \leq d \leq 10$ cm. The upper limit of 10 cm was chosen to approximate the largest particle size blown by the rocket exhaust, but the integral is not sensitive to that value. This function was derived by fitting to a concatenation of submicron data from Park et al. (2008) with data from JSC-1A weighted such that 8% of the mass is finer than 10 microns. With these inputs we can now calculate the impacting flux per Eq. 4. The result is $\psi_{\text{impact}} = 612$ mg/cm²/s, so during the 11.5 ms duration passage through the ejecta cone the spacecraft will experience a total of 7 mg/cm² impacting soil at hypervelocity. The number of discrete soil particles impacting the surface is

$$N_{\text{Impact}} = \frac{\dot{M}\Delta t}{2\,\pi\,R^2\,\Delta\theta} \int_{d_{\min}}^{d_{\max}} \left(\frac{P(d)}{\rho\,\pi d^3/6}\right)\left(\frac{0.940\,v(d) + v_{\text{LLO}}}{v(d)}\right) \mathrm{d}d \tag{6}$$

where $d_{\min} = 0.019\ \mu$m for the smallest size of lunar soil particles per Park et al. (2008). This predicts $N_{\text{Impact}} = 256$ million impacts/m², mostly very small. Using the aluminum damage equations of Lambert (1997), Figure 8 shows the number of impactors at each range of damage. Using the glass spallation equation of Jiyun, et al. (2010), we calculate 4% of the surface of exposed glass will be spalled.

**CONCLUSION**

The damage to the Lunar Orbital Gateway is slight, but a spacecraft in Low Lunar Orbit may suffer extensive damage if the timing of its orbit puts it in the path of the ejecta cone from a large lunar lander. The damage will be worse for landers with greater mass or fewer engines or if they continue firing the engines closer than 1.5 m to the surface. Future work will better constrain these estimates.

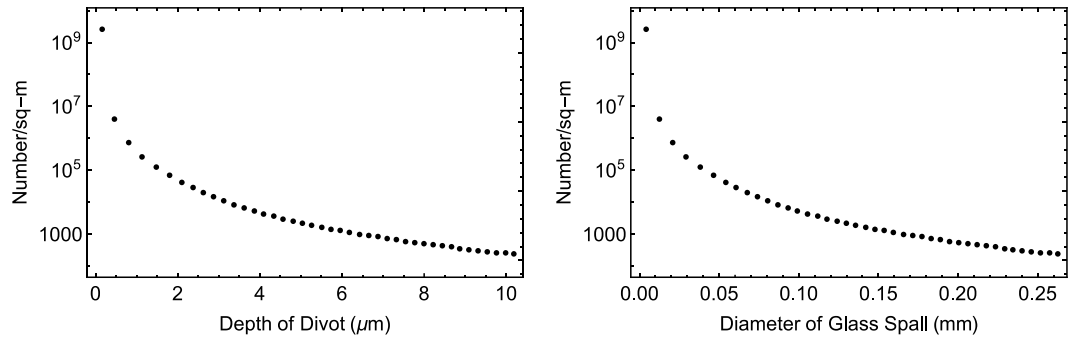

**Figure 8. Damage in LLO. Left: for aluminum target material, number of divots/m² for each depth of divot. Right: for glass target material, number of spallations/m² for each diameter of spallation.**